\documentclass[journal,12pt,onecolumn]{IEEEtran}
\usepackage{epsf,epsfig,eepic,epic}
\usepackage{fancybox}
\usepackage{amsmath,amssymb,amsbsy}
\usepackage{euscript,bm}
\usepackage{theorem}
\usepackage{array,longtable}
\usepackage{times}
\usepackage[gen]{eurosym}
\usepackage{graphics,graphicx}
\usepackage{fancyhdr}
\usepackage{cite}
\usepackage{subfigure}
\usepackage{placeins}
\usepackage{float}
\usepackage{lmodern}
\usepackage{aeguill}
\usepackage[final]{pdfpages} 
\usepackage{verbatim}
\usepackage[Lenny]{fncychap}
\usepackage{url} 
\usepackage{tocbibind}
\usepackage[squaren,Gray]{SIunits}
\usepackage{dsfont}
\usepackage{hyperref}
\usepackage{multirow}
\usepackage{comment}
\usepackage{algorithm,algpseudocode}
\usepackage{geometry}
\usepackage{booktabs}
\usepackage{tabularx}
\usepackage{makecell}
\usepackage{stmaryrd}
\usepackage{lineno}
\usepackage{etoolbox} 
\usepackage{lipsum}   

\makeatletter
\let\old@eqnnum\@eqnnum
\patchcmd{\@eqnnum}{\hbox}{\hb@xt@.01\p@}{}{}
\makeatother

\begin{document}

\title{CAZAC sequence generation of any length with iterative projection onto unit circle: principle and first results}

\author{Karine Amis,~\IEEEmembership{Member,~IEEE}, Eloi Boutillon and Emmanuel~Boutillon,~\IEEEmembership{Senior Member,~IEEE}
% Emmanuel - ORCID : 0000 0003 2124 0786
% Eloi - ORCID : 0009-0006-6698-8308
% Karine - ORCID : 0000-0002-3130-7772
\thanks{Karine Amis is with IMT-Atlantique, Lab-STICC UMR 6285, 29238 Brest, France (e-mail: karine.amis@imt-atlantique.fr).}
\thanks{Eloi Boutillon contributes to this work in his personal time (e-mail: eloi.boutillon@outlook.fr).}
\thanks{Emmanuel Boutillon is with Université Bretagne Sud, Lab-STICC UMR 6285, France (e-mail: emmanuel.boutillon@univ-ubs.fr).}}

\maketitle

\begin{abstract}
Constant amplitude zero-autocorrelation (CAZAC) sequences are mainly used for synchronization in communication and radar applications. The state-of-the-art proposes analytical derivation of specific families whose major limitation comes from the alphabet which only represents a fraction of the whole, the longer the sequences, the smaller the fraction. The objective of the paper is threefold, first to present the construction of constant amplitude zero-circular autocorrelation sequences of any length using iterative projection onto Unit Circle (IPUC) algorithm. This algorithm allows, from any random seed, to generate a near-CAZAC sequence. Then, focusing on length-8 sequences, we propose a classification of the IPUC output with an analytical expression of a representative for each identified equivalence class. Finally, the IPUC is applied within a simulated-annealing process to generate near-CAZAC sequences suitable for radar applications with optimized ratio between first and second lobes of the non-circular autocorrelation function.
\end{abstract}
\IEEEoverridecommandlockouts

\IEEEpeerreviewmaketitle

\section{Introduction}

\IEEEPARstart{T}{he} literature on constant amplitude zero autocorrelation (CAZAC) sequences is quite abundant. An $n$-periodic sequence $\bm{x}$ is said to be a CAZAC sequence if and only if all elements of $\bm{x}$ have modulus $1$ and if the circular autocorrelation function is zero except in $0$ where it is equal to $n$. CAZAC sequences are used in many communication systems (mainly for synchronization purposes) such as 3GPP. An accurate and detailed overview of CAZAC sequences and their properties can be found in \cite{Benedetto,Benedetto2019}. Among known sequences, one can cite the Zadoff-Chu sequences\cite{Chu_sequence, Chu_sequence2, ZC_3} and their extension by Popovic in \cite{chirp,Popovic2010}. P4\cite{Kretschmer} sequences as well as Wiener\cite{Wiener} sequences are other well-known CAZAC sequences. Björck also proposed two families in \cite{Bjorck} with restrictions on the length. Table~\ref{tab:CAZAC} resumes the different types of known CAZAC sequences. 

However these families only cover a fraction of the available CAZAC sequences (the longer the length, the smaller the fraction). In \cite{10549571}, an algorithm referred to as polar-optimized is proposed to generate CAZAC sequences from an optimization problem using the ZAC criterion as the function cost. However the simulation time reported in \cite{10549571} is rather long and the output solutions have not been analysed.

\begin{table*}
\centering
\renewcommand{\arraystretch}{1} % Increase row spacing for better readability
\caption{Classification of state-of-the-art CAZAC sequences}
\label{tab:CAZAC}
\begin{tabular}{p{0.12\textwidth}p{0.12\textwidth}p{0.66\textwidth}}
\toprule
\textbf{Type} & \textbf{Values of $n \in \mathbb{N}$} & \textbf{Phase expression $\theta(k)$ of the $k^{th}$ terms $x(k) = e^{i\theta(k)}$} \\
\midrule

Zadoff-Chu \cite{Chu_sequence, Chu_sequence2, ZC_3}& $n>1$ & $\theta(k) = -\pi\frac{ u k(k+c+2q)}{n}$ with $u,q \in \mathbb{N}$, $u,q <n,\: \gcd(n,u)=1$ and $c = n \bmod 2$. \\
&&\\
Popovic \cite{chirp,Popovic2010}& $n = m^2t$ & $\theta(k) = zc(k) + 2\pi w( k \bmod m)$ with $zc$ a length-$n$ Zadoff-Chu sequence and $w$ a length-$m$ sequence with real-valued components in $[0,1]$.\\
&&\\
Wiener \cite{Wiener,Benedetto}
 & $n>1$ & $\theta(k) = 2\pi\frac{mk^2}{p}$ with $p=n$ if $n$ is odd and $p=2n$ if $n$ is even, $\gcd(p,m) = 1$.\\
 &&\\
Lewis-Kretschmer (P4) \cite{Kretschmer,Benedetto} & $n >1$ & $\theta(k) = \pi \frac{k(k-n)}{n}$\\
 Björck I \cite{Bjorck} & \makecell[l]{$n$ prime \\ $n\bmod4=1$} & $\theta(k) = \left(\frac{k}{n}\right)_L\arccos\left(\frac{1}{1+\sqrt{n}}\right) $ with $\left(\frac{k}{n}\right)_L$ the Legendre symbol equal to $1$ if $\exists u \neq 0$ so that $u^2 \mod n = k$, 0 if $k=0$, and $-1$ otherwise. \\
&&\\
Björck II \cite{Bjorck} & \makecell[l]{$n$ prime \\ $n\bmod4=3$} & $\theta(k) = \begin{cases} \arccos\left(\frac{1-n}{1+n}\right) \: \text{if } \left(\frac{k}{n}\right)_L= -1 \\
       0 \text{ otherwise} \end{cases}$ \\

\bottomrule
\end{tabular}
\end{table*}

In this paper, we are adding three contributions to this literature: first, the proposition of a new algorithm called iterative projection onto Unit Circle (IPUC) to rapidly generate near-CAZAC sequences, second, the analysis of the structure of some solutions found for $n=8$ allows us to propose a classification into 4 types with explicit analytical expression. Zero-degree-of-freedom sequences in particular have  been identified and divided into equivalence classes. 
Finally, we explain how to take profit of the IPUC algorithm to construct a near-CAZAC sequence with an additional property beneficial for radar applications, i.e., CAZAC sequences that minimize the maximum second lobe of the non-circular autocorrelation function \cite{10549571}. 

This paper is organized as follows. The second section is a reminder of most CAZAC sequence properties among which one of them is used in the third section to derive the iterative projection onto Unit Circle (IPUC)  algorithm. The fourth section is dedicated to the classification into four types of CAZAC sequences for $n=8$. 

\section{Reminder about CAZAC sequences properties for generation criteria}
Let $\bm{x} = (x(0), x(1), \ldots, x(n-1))$ a length-$n$ sequence. $\bm{x}$ is said to be constant amplitude (CA) if there exists $K\geq 0$ such that $|x(k)|=K$ whatever $k \in \{0,\dots,n-1\}$. In the remainder of this paper we consider $K=1$ and $x(k)=e^{i\theta(k)}$ with $\theta(0) = 0$. 

\subsection{Properties and design criteria}\label{sec:cazac_properties}
We define the length-$n$ circular autocorrelation vector $\bm{R_x}$ by: 
\begin{equation}
    \bm{R_x}(k) = \sum_{\ell=0}^{n-1}x(\ell)x^{\star}\left(\left(\ell-k\right)\mod{n}\right),
    \label{eq:circ_autocorr}
\end{equation}

\noindent with $x^{\star}$ the complex conjugate of $x$.
Then $\bm{x}$ is said to be zero autocorrelation (ZAC) if $\bm{R_x} = \begin{pmatrix}
    n & \bm{0}_{n-1}
\end{pmatrix}$. 
    
Let us denote by $\mathcal{F}$ the discrete Fourier transform (discrete FT) operation defined in Table~\ref{table:transformations}. Then the CA sequence $\bm{x}$ is ZAC if and only if $\bm{X}=\mathcal{F}\left(\bm{x}\right)$ is CA (see \cite{Benedetto} for the proposition reminder and the proof in Appendix \ref{appendix:cazac_prop}). This criterion is the key principle of the iterative algorithm proposed in Section \ref{sec:ipuc}. to generate near-CAZAC sequences. The reader can refer to the Appendix \ref{appendix:cazac_prop} for the demonstration and further details.

Taking into account of the above, we obtain another criterion for CAZAC generation also proved in Appendix \ref{appendix:cazac_prop}. A CA sequence $\bm{x}=e^{i\bm{\theta}}$ is ZAC if and only if 
\begin{equation}
\sum_{\ell=0}^{n-1}e^{i\left(\theta(\ell)-\theta(\left(\ell-k\right)\mod{n})\right)}=0, \; \forall k \neq 0  
\label{eq:analytical_criterion}
\end{equation}
Let us mention that this criterion can be found in an equivalent form with different demonstration elements in \cite{10549571}. In \cite{10549571} a polar-based generation algorithm using this criterion as a cost function to minimize is proposed and applied to derive different CAZAC sequences with length $n\in\{10,14,15\}$. This algorithm is to be compared to the one proposed in Section \ref{sec:ipuc}. In this paper, we will use this criterion in Section \ref{sec:analytic_class} to analytically derive CAZAC sequences for $n=8$.

 An important property, a consequence of the above, is: $\bm{x}$ is CAZAC if and only if $\bm{X}$ is CAZAC. To a greater extent, from an elementary CAZAC sequence, one can derive other CAZAC sequences using the CAZAC-invariant transformations reported in Table~\ref{table:transformations}, alone or combined \cite{Benedetto2019}.
 \begin{table}
     \centering
     \caption{List of CAZAC-invariant transformations. Note that $\mathbb{P}(n)$ is the set of natural numbers less than $n$ and prime with $n$}.
     \begin{tabular}{r|l|l} \hline
        Type  & Function & $(y(k))_{k \in \llbracket 0, n-1  \rrbracket}$ \\ \hline
        Rotation &$\bm{y} = R_{\theta \in \mathbb{R}}(\bm{x})$ &  $e^{i\theta}x(k)$ \\
        Translation& $\bm{y} = T_{r \in \mathbb{N}} (\bm{x})$ &  $x\left((k+r) \bmod n\right)$ \\
        Decimation &$\bm{y} = D_{d \in \mathbb{P}(n)}(\bm{x})$ &  $x(dk \bmod n)$ \\
        Modulation &$\bm{y} = M_{m \in \mathbb{N}}(\bm{x})$ &   $e^{i\frac{2\pi  mk}{n}}x(k)$ \\
        Conjugaison &$\bm{y} = C_{a\in \{0,1\}}(\bm{x})$ &  $\bm{x}$ if $a = 0$, $\bm{x}^\star$ if $a=1$\\
        Discrete FT &$\bm{y} = \mathcal{F}(\bm{x})$ &  $\frac{1}{\sqrt{n}}\sum_{l =0}^{n-1}e^{-i\frac{2\pi lk}{n}}x(l)$ \\ \hline
     \end{tabular}
     \label{table:transformations}
\end{table}
 
\textbf{Definition} Two CAZAC sequences $\bm{x}$ and $\bm{y}$ are said to belong to the same equivalence class if and only if there exists a combination of CAZAC-invariant transformations to derive $\bm{y}$ from $\bm{x}$.  

For example, for $n = 8$, Tables~\ref{tab:ZC},~\ref{tab:PP},~\ref{tab:P4} and~\ref{tab:Wiener} provide a representative of each class for Zadoff-Chu, Popovic, Lewis-Kretschmer (P4) and Wiener sequences respectively. We observe that Zadoff-Chu, P4 and Wiener sequences can be obtained from a Popovic sequence for a given $\theta_1$ (either from the representative or its conjugate).
\begin{table}
  \centering
  \caption{Zadoff-Chu sequences for $n=8$}
  \begin{tabular}{rrrrrrrr}
    \hline
    1&$e^{i\frac{\pi}{8}}$&1&$-ie^{i\frac{\pi}{8}}$&-1&$e^{i\frac{\pi}{8}}$&-1&$-ie^{i\frac{\pi}{8}}$\\
    1&$e^{i\frac{\pi}{8}}$&$i$&$-e^{i\frac{\pi}{8}}$&1&$-e^{i\frac{\pi}{8}}$&$i$&$e^{i\frac{\pi}{8}}$\\
    1&$e^{i\frac{\pi}{8}}$&-1&$ie^{i\frac{\pi}{8}}$&-1&$e^{i\frac{\pi}{8}}$&1&$ie^{i\frac{\pi}{8}}$\\
    1&$e^{i\frac{\pi}{8}}$&$-i$&$e^{i\frac{\pi}{8}}$&1&$-e^{i\frac{\pi}{8}}$&$-i$&$-e^{i\frac{\pi}{8}}$\\
    \hline
    \end{tabular} 
    \label{tab:ZC}
    \end{table}
\begin{table}
  \centering
  \caption{Generic Popovic set $\mathcal{P}$ for $n=8$ ($\theta \in \mathbb{R}$)}
  \begin{tabular}{rrrrrrrr}
    \hline
    1&$e^{i\theta}$&1&$-ie^{i\theta}$&-1&$e^{i\theta}$&-1&$-ie^{i\theta}$\\
    1&$e^{i\theta}$&$i$&$-e^{i\theta}$&1&$-e^{i\theta}$&$i$&$e^{i\theta}$\\
    1&$e^{i\theta}$&-1&$ie^{i\theta}$&-1&$e^{i\theta}$&1&$ie^{i\theta}$\\
    1&$e^{i\theta}$&$-i$&$e^{i\theta}$&1&$-e^{i\theta}$&$-i$&$-e^{i\theta}$\\
    \hline
    \end{tabular} 
    \label{tab:PP}
    \end{table}
\begin{table}
  \centering
  \caption{Lewis-Kretschmer (P4) set for $n=8$ }
  \begin{tabular}{rrrrrrrr}
    \hline
    1&$-e^{i\frac{\pi}{8}}$&$i$&$e^{i\frac{\pi}{8}}$&1&$e^{i\frac{\pi}{8}}$&$i$&$-ie^{i\frac{\pi}{8}}$\\
    \hline
    \end{tabular} 
    \label{tab:P4}
    \end{table}
    \begin{table}
  \centering
  \caption{Wiener set  for $n=8$}
  \begin{tabular}{rrrrrrrr}
    \hline
    1&$e^{i\frac{\pi}{8}}$&$i$&$-e^{i\frac{\pi}{8}}$&1&$-e^{i\frac{\pi}{8}}$&$i$&$e^{i\frac{\pi}{8}}$\\
    1&$e^{i\frac{3\pi}{8}}$&$-i$&$-e^{i\frac{3\pi}{8}}$&1&$-e^{i\frac{3\pi}{8}}$&$-i$&$e^{i\frac{3\pi}{8}}$\\
    1&$-e^{-i\frac{\pi}{8}}$&$-i$&$e^{-i\frac{\pi}{8}}$&1&$e^{-i\frac{\pi}{8}}$&$-i$&$-e^{-i\frac{\pi}{8}}$\\
    1&$-e^{-i\frac{3\pi}{8}}$&$i$&$e^{-i\frac{3\pi}{8}}$&1&$e^{-i\frac{3\pi}{8}}$&$i$&$-e^{-i\frac{3\pi}{8}}$\\
    \hline
    \end{tabular} 
    \label{tab:Wiener}
    \end{table}
\section{Iterative algorithm to generate "almost" CAZAC sequences}
\label{sec:ipuc}
This section presents an iterative algorithm to generate a CAZAC sequence of any length. It is based on an iterative projection onto the unit circle (IPUC). The algorithm is first detailed and illustrated through some simulation results before being extended to provide CAZAC sequences appropriate for radar applications.
\subsection{Description of the IPUC algorithm}
Let us define the discrepancy function $D(\bm{x})$ that measures how far a sequence $\bm{x}$ is from being a true CAZAC sequence. The function $D(\bm{x})$ is the sum of two terms $D(\bm{x}) = D_{CA}(\bm{x}) + D_{ZAC}(\bm{x})$. The first term $D_{CA}(\bm{x})$ indicates the maximum absolute discrepancy between the modulus of $\bm{x}$ and the modulus of a true CA sequence i.e., 
\begin{equation} 
    D_{CA}(\bm{x}) = \max_{0 \leq k <n} \left( \left||x(k)| - 1\right|\right).
\end{equation}

The second term, $D_{ZAC}(\bm{x})$,  indicates the maximum absolute discrepancy between the circular autocorrelation function of $\bm{x}$ and the ZAC function of a true CAZAC sequence, i.e.

\begin{equation}
    D_{ZAC}(\bm{x}) = \max_{0<k<n} \left( \left| \bm{R}_{\bm{x}}(k)\right|\right).
\end{equation}

\begin{figure}[!t]
    \centering
    \includegraphics[width=0.8\columnwidth]{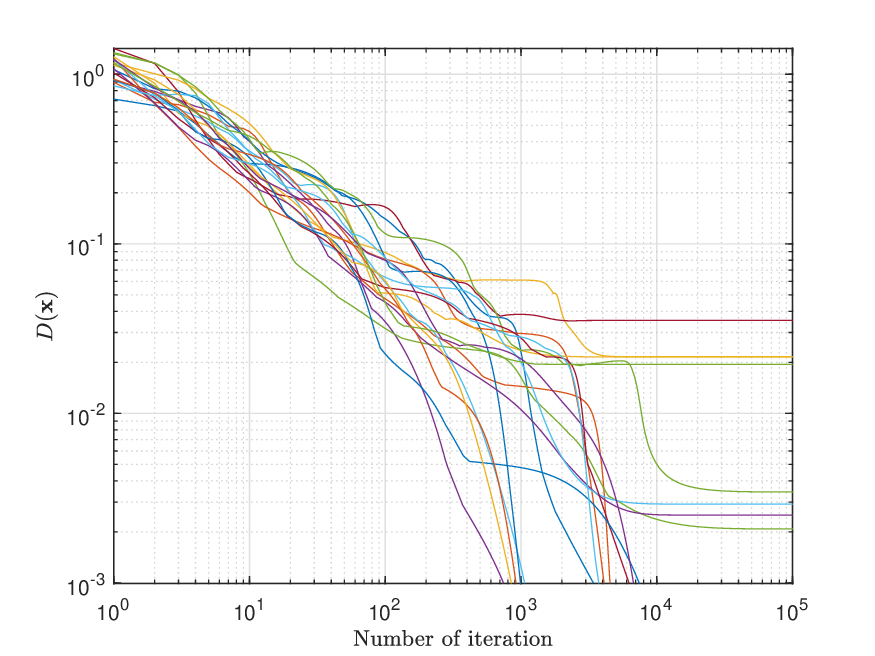}
    \caption{Example of trajectories of the discrepancy function during the generation process of 20 CAZAC sequences of length $n=50$.}
    \label{fig:trajectory}
\end{figure}

Note that if $D(\bm{x})=0$, then $\bm{x}$ is a CAZAC sequence. The proposed iterative algorithm starts from a random CA sequence $\bm{X}$ in the frequency domain, then goes back and forth in the time/frequency domain, each time projecting the obtained constellation onto the unit circle, i.e., a CA sequence in the time domain and a CA sequence in the frequency domain, as shown in Algorithm \ref{algo:projective_algorithm}. Figure \ref{fig:trajectory} shows 20 trajectories of the discrepancy function ($y$-axis) as a function of the iteration number (x-axis) given by the IPUC algorithm starting from 20 random seeds with length $n=50$.

In this example, 5 sequences with a discrepancy lower than $10^{-3}$ are obtained in fewer than $10^{3}$ iterations, 7 additional sequences are obtained in fewer than $10^{4}$ iterations, and finally, the remaining 8 sequences get a discrepancy frozen above $10^{-3}$ after $10^{5}$ iterations. In practice, to generate CAZAC sequences efficiently, if the convergence is not fast enough, it is more efficient to restart from a new seed than to run (or waste) a large number of additional iterations for a potential future convergence. Although very interesting, the analysis of this iterative function is beyond the scope of this paper. In practice, the algorithm can determine a CAZAC sequence with $D(\bm{x}) < 10^{-3}$ in less than a few seconds for any value of $n$ less than 1000, and in around 30 seconds for $n=10,000$.

\begin{algorithm}
\textbf{Input} A random CA sequence $\bm{X}$ of length $n$, maximum discrepancy target equal to $\epsilon$ \\
\textbf{Output} A near-CAZAC-sequence of length $n$ and a maximum discrepancy of $\epsilon$.

\begin{algorithmic}
\State   $\bm{x} = \mathcal{F}^{-1}(\bm{X})$;

\While {$D(\bm{x}) > \epsilon$} 
    \State $\bar{\bm{x}} = \left(\frac{x(0)}{|x(0)|},\frac{x(1)}{|x(1)|}, \ldots, \frac{x(n-1)}{|x(n-1)|}\right)$;
    \State $\bm{X} = \mathcal{F}(\bar{\bm{x}})$;
    \State $\bar{\bm{X}} = \left(\frac{X(0)}{|X(0)|},\frac{X(1)}{|X(1)|}, \ldots, \frac{X(n-1)}{|X(n-1)|}\right)$
    \State $\bm{x} = \mathcal{F}^{-1}(\bar{\bm{X}})$;
\EndWhile
\State \textbf{Return} $\bm{x}$;
\end{algorithmic}
\caption{Iterative projection algorithm to generate a CAZAC-sequence.}
\label{algo:projective_algorithm}
\end{algorithm}

As previously mentioned, the efficiency of Algorithm \ref{algo:projective_algorithm} can be significantly improved by introducing additional control to reset the iterative process with a new seed when the decreasing speed of $D(\bm{x})$ becomes too slow. Appendix \ref{annex:result} gives examples of generated CAZAC sequences of length 8 with the corresponding type of sequence.

\subsection{Extension for additional constraint on the non-circular autocorrelation}

The IPUC algorithm can also be included in a higher-level optimization process to generate CAZAC sequences with auxiliary properties. Let us illustrate its extension in the context of radar applications. The non-circular autocorrelation function $\bm{R}^\text{nc}_x$ of a sequence $\bm{x}$ is defined by
\begin{equation}
 \bm{R}^\text{nc}_{\bm{x}}(\tau) =  \sum_{k=0}^{n-1-|\tau|} x(k)x^\star(k+|\tau|).
 \label{eq:nc_auto_corr}
\end{equation}
Let us mention the difference with the circular autocorrelation definition in (\ref{eq:circ_autocorr}) where the sum involves $n$ terms (modulo-$n$ operation). For radar applications, it could be important to increase the energy ratio $\rho $ between the first and the second lobes of $\bm{R}^\text{nc}_{\bm{x}}$, which is defined in decibels as follows
\begin{equation}
\rho_\text{dB}  = 10 \log_{10}\left( \frac{|\bm{R}^\text{nc}_{\bm{x}}(0)|^2}{\underset{0<\tau< n}{\max} \{|\bm{R}^\text{nc}_{\bm{x}}(\tau)|^2 \}}\right) \: \: \text{(dB)} 
\label{eq:rho}    
\end{equation}
Since the first lobe has a constant energy $|{R}^\text{nc}_x(0)|^2 = n^2$ for CAZAC sequences, the optimization problem becomes 
\begin{equation}
 \hat{\bm{x}} = \arg \min_{\bm{x} \in \mathcal{C}_n} \{ \max_{0<\tau <n} |\bm{R}^\text{nc}_{\bm{x}}(\tau)|^2 \},
 \label{eq:crit_nc_auto_corr}
\end{equation}
\noindent with $\mathcal{C}_n$ the set of CAZAC sequences of length $n$.

\begin{figure}[!t]
    \centering
    \includegraphics[ width=0.8\columnwidth]{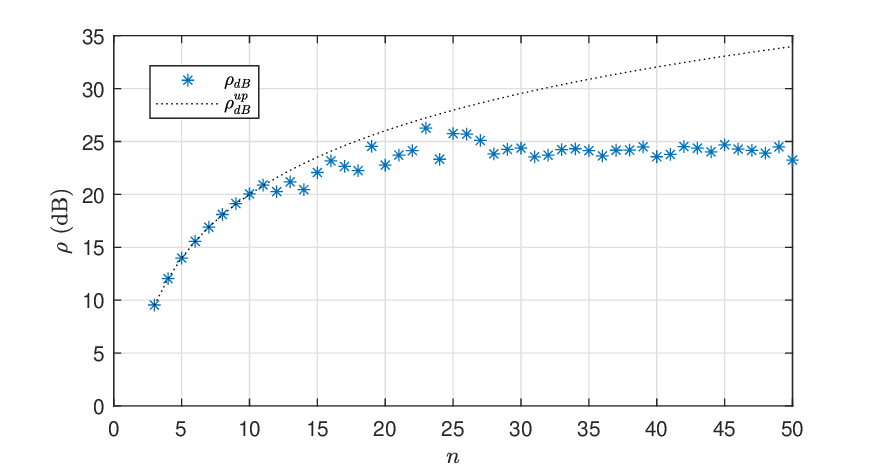}
    \caption{Energy ratio between first and second lobes of the non-circular  autocorrelation function of CAZAC sequences for length $n$ varying from 2 to 50.}
    \label{fig:energy_ratio}
\end{figure}

The implementation of the CAZAC generation function in a simulated-annealing process allows one to generate CAZAC sequences optimized to minimize the criterion in (\ref{eq:crit_nc_auto_corr}).  The efficiency of such a procedure is illustrated in figure \ref{fig:energy_ratio} where the energy ratio $\rho_{\text{dB}}$ between the first and the second lobes of $\bm{R}^\text{nc}_{\bm{x}}$  for output CAZAC sequences is plotted as a function of the length $n$ ranging from 2 to 50. 

Note that for $\tau = n-1$, $|\sum_{k=0}^{n-1-\tau} x(k)x^\star(k+\tau)|^2 = |x(0)x^\star(n-1)|^2 = 1$, thus, $\rho_{\text{dB}}$ is upper-bounded by $\rho^{\text{u}}_{\text{dB}} = 10\log_{10}(n^2)$. An example of a phase sequence obtained for $n=23$ is $\theta = \frac{2\pi}{23}\times$( 0,   16.0884,   12.7028,    8.9221,    6.9862,    1.1362,   12.7345,    2.3399,   22.8821,   13.8704,    1.5708,   14.8121,   22.5770,    9.8769,  16.8806,   17.3456,    2.7453,   12.1426,   15.9850,   15.9248,   17.7010,   19.0881,    1.0068). It generates a CAZAC sequence $\bm{x} = e^{i\theta}$ with $\rho_{\text{dB}} = 26.25$ dB. This value is close to the upper bound $\rho^{\text{u}}_{\text{dB}} = 27.23$ dB.

\section{Analytical determination of CAZAC sequences}\label{sec:analytic_class}

In this section, we caracterize the sets of CAZAC sequences for $n=4$ and $n=8$. For $n=8$, we start from a given format observed at the IPUC algorithm output (Algorithm \ref{algo:projective_algorithm}) and then we apply the criterion given in  (\ref{eq:analytical_criterion}) to derive the corresponding generic family of sequences.

\subsection{CAZAC sequences with \texorpdfstring{$n=4$}{n = 4}}

Let us denote the CA sequence by $\bm{x}=\left(
    1,e^{i\theta},e^{i\beta},e^{i\gamma}
\right)$.
Then the ZAC criterion in (\ref{eq:analytical_criterion}) yields the following simplified system
\begin{equation}
    \left\{
    \begin{array}{lll}
    e^{-i\gamma}+e^{i\theta}+e^{i(\beta-\theta)}+e^{i(\gamma-\beta)}&=&0\\
    \cos(\beta)+\cos(\theta-\gamma)&=&0
    \end{array}
    \right.
\end{equation}
whose solution is given by
\begin{equation}
    \left\{
    \begin{array}{lll}
    \beta&=&\theta-\gamma+\pi\pmod{2\pi}\\
    \gamma&=&\theta \pmod{\pi}
    \end{array}
    \right.
\end{equation}
More precisely, the set of CAZAC sequences for $n=4$ is $\mathcal{S}=\{\left(
    1,e^{i\theta},-1,e^{i\theta}
\right),\theta \in \mathbb{R}\}\cup \{\left(
    1,e^{i\theta},1,-e^{i\theta}
\right),\theta \in \mathbb{R}\}$
One can observe that the two resulting subsets are orthogonal and that they correspond to Popovic sequences for $n = 4$ with $\theta$ as the degree of freedom. 

For $n=4$, the IPUC algorithm always generates a CAZAC sequence that belongs to one of these two equivalence classes. 

\subsection{CAZAC sequences with \texorpdfstring{$n=8$}{n = 8}}

This section is dedicated to the classification into four types and analytical derivation of CAZAC sequences in the case of $n=8$. We first give the generic criterion before detailing each type. For the ease of formulation, in the rest of the paper, we define $\theta_{\ell p}=\theta_\ell-\theta_p$ such that $\theta_{\ell 0} = \theta_\ell$. 
\subsubsection{ZAC criterion formulation for \texorpdfstring{$n=8$}{n = 8}}
Let $\bm{x}=\left(
1,e^{i\theta_1},e^{i\theta_2},e^{i\theta_3},e^{i\theta_4},e^{i\theta_5},e^{i\theta_6},e^{i\theta_7}
\right)$ be a CAZAC sequence.
Then the ZAC criterion yields the following simplified system to be solved:

\begin{equation}
    \left\{
    \begin{array}{l}
    e^{-i\theta_{70}}+e^{i\theta_{10}}+e^{i\theta_{21}}+e^{i\theta_{32}}+e^{i\theta_{43}}+e^{i\theta_{54}}
    \\
    +e^{i\theta_{65}}+e^{i\theta_{76}}=0\\
    
    e^{-i\theta_{60}}+e^{i\theta_{17}}+e^{i\theta_{20}}+e^{i\theta_{31}}+e^{i\theta_{42}}+e^{i\theta_{53}}\\
    +e^{i\theta_{64}}+e^{i\theta_{75}}=0\\
    e^{-i\theta_{50}}+e^{i\theta_{16}}+e^{i\theta_{27}}+e^{i\theta_{30}}+e^{i\theta_{41}}+e^{i\theta_{52}}\\
    +e^{i\theta_{63}}+e^{i\theta_{74}}=0\\
    \cos(\theta_{40})+\cos(\theta_{15})+\cos(\theta_{26})+\cos(\theta_{37})=0
    \end{array}
    \right.
    \label{eq:non_linear_system}
\end{equation}

\subsubsection{Analysis of solution formats at the IPUC algorithm output}

The IPUC algorithm described in Algorithm \ref{algo:projective_algorithm} allowed us to generate tens of thousands of length-8 CAZAC sequences. Given the output sequence $\bm{x}=e^{i\bm{\theta}}$, a rotation $R_{-\theta(1)}$ is applied (see Table~\ref{table:transformations} for the definition) so that the resulting CAZAC sequence starts by 1. Let $\mathcal{S}$ be the set of generated sequences.

\begin{figure*}[!t]
    \centering
     \includegraphics[ width=\textwidth]{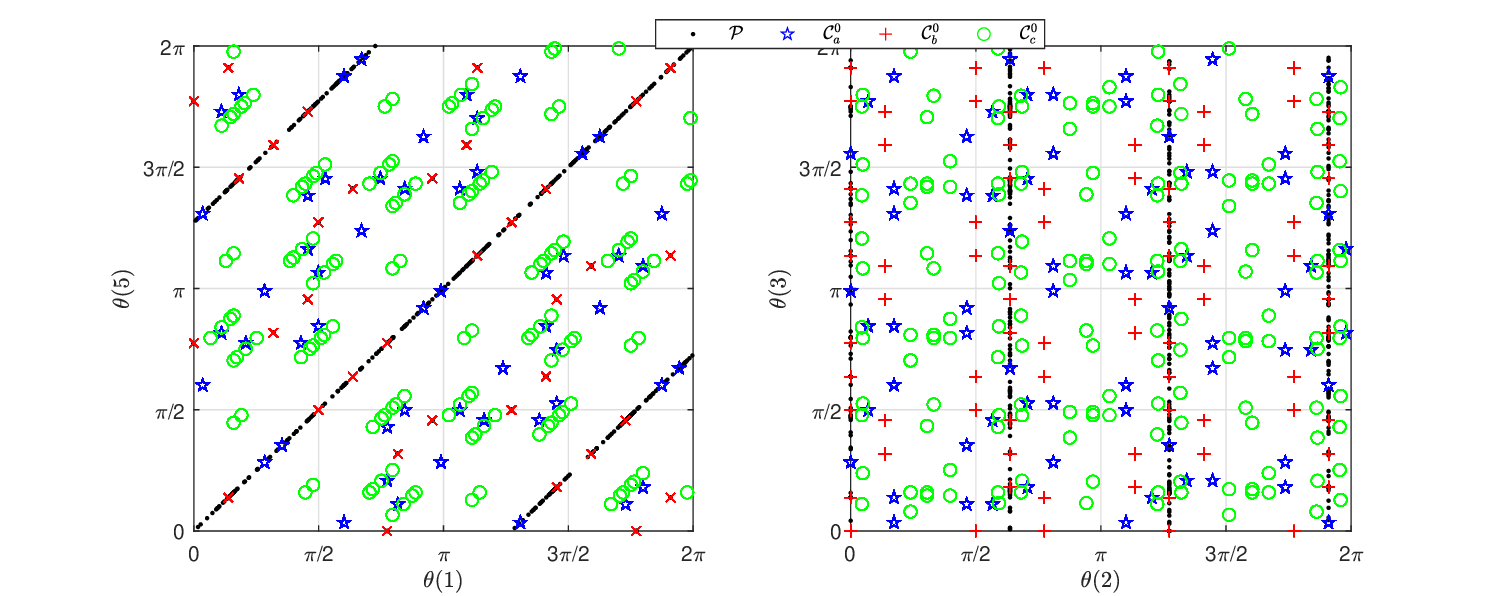}
    \caption{Highlighting the existing structure of length-8 CAZAC sequences. Plot of $\theta(1)$ (resp. $\theta(2)$) as a function of $\theta(5)$ (resp. $\theta(3)$) from 10,000 generated near-CAZAC sequences. The class of each CAZAC sequence is identified thanks to a specific symbol (black point for $\mathcal{P}$), blue pentagram for $\mathcal{C}^0_a$, red plus sign for $\mathcal{C}_b^0$ and green circle for $\mathcal{C}_c^0$.}
    \label{fig:graphe_phase}
\end{figure*}

Figure \ref{fig:graphe_phase} shows the correlation between the components of sequences $\bm{x} \in \mathcal{S}$. For each sequence, the component $\theta(5)$ is plotted as a function of $\theta(1)$ on the left side while $\theta(3)$ is plotted as a function of $\theta(2)$ on the right side. The figure is generated from $10^4$ CAZAC sequences obtained by the IPUC algorithms. Note that among the 10,000 generated sequences, only 681 are of type $\bm{x} \in \mathcal{P}$ (around 7 \% of the resulting sequences). The relation between the phase components of those sequences draws the lines in figure \ref{fig:graphe_phase}. The isolated points correspond to the new types of sequences. These two graphs show that the sequences do not cover the whole space and that some structure exists. This is what motivated us to establish a classification of the solutions obtained by the IPUC algorithm.

It appears that all generated sequences belong to one of the four following families (defined through the format of their representatives):
\begin{align}
    \mathcal{F}_1 &= \{(1,e^{i\theta_1},e^{i\theta_2},e^{i\theta_3},e^{i\theta_4},e^{i\theta_3},e^{i\theta_2},e^{i\theta_1})\}
,\\
    \mathcal{F}_2 &= \{(1,e^{i\theta_1},e^{i\theta_2},e^{i\theta_1},e^{i\theta_4},e^{i\theta_5},e^{i\theta_2},e^{i\theta_5})\},\\
    \mathcal{F}_3 &= \{(1,e^{i\theta_1},e^{i\theta_2},e^{i\theta_3},e^{i\theta_4},e^{i\theta_1},e^{i\theta_6},e^{i\theta_3})\},\\
    \mathcal{F}_4 &= \{(1,e^{i\frac{\pi}{8}},e^{i\theta_2},e^{i\theta_3},e^{i\theta_4},e^{i f(\theta_4)},e^{i g(\theta_3)},e^{i h(\theta_2)}\},
\end{align}
where $\theta_i$, $i = 1,2, \ldots, 6$ are real values and $f(x) = x -\frac{7\pi}{8}$, $g(x) = x -\frac{5\pi}{8}$ and $h(x) = x -\frac{3\pi}{8}$. Each of the four forms will now be analytically derived through the non-linear system resolution written in (\ref{eq:non_linear_system}) after imposing the format as an hypothesis.

Let us consider the angle sequence $\bm{\theta}=(\theta_0,\theta_1,\theta_2,\theta_3,\theta_4,\theta_5,\theta_6,\theta_7)$. Then the three versions of this sequence following the decimation operation are:
\begin{eqnarray*}
    \bm{\theta}_{m=3}&=&(\theta_0,\theta_3,\theta_6,\theta_1,\theta_4,\theta_7,\theta_2,\theta_5)\\
     \bm{\theta}_{m=5}&=&(\theta_0,\theta_5,\theta_2,\theta_7,\theta_4,\theta_1,\theta_6,\theta_3)\\
    \bm{\theta}_{m=7}&=&(\theta_0,\theta_7,\theta_6,\theta_5,\theta_4,\theta_3,\theta_2,\theta_1)
\end{eqnarray*}
We observe that $\mathcal{F}_1$ is the set of sequences invariant by decimation $m=7$ ($\bm{\theta}_{m=7}=\bm{\theta}$), $\mathcal{F}_2$ the set of sequences invariant by decimation $m=3$ ($\bm{\theta}_{m=3}=\bm{\theta}$) and $\mathcal{F}_3$ is the set of sequences invariant by decimation $m=5$ ($\bm{\theta}_{m=5}=\bm{\theta}$).

\subsubsection{First generic form}
The first form has four parameters $\{\theta_1,\theta_2,\theta_3,\theta_4\}$. Combining the structure of $\mathcal{F}_1$ to (\ref{eq:non_linear_system}) gives the following non-linear system:
\begin{equation*}
    \left\{
    \begin{array}{l}
    \cos(\theta_{10})+\cos(\theta_{21})+\cos(\theta_{32})+\cos(\theta_{43})=0\\
    1+\cos(\theta_{20})+\cos(\theta_{31})+\cos(\theta_{42})=0\\
    \cos(\theta_{30})+\cos(\theta_{21})+\cos(\theta_{32})+\cos(\theta_{41})=0\\
    1+\cos(\theta_{40})+2\cos(\theta_{31})=0
    \end{array}
    \right.
\end{equation*}

An obvious solution to $1+\cos(\theta_{40})+2\cos(\theta_{31})=0$ yields 
\begin{equation}
    \left\{
    \begin{array}{l}
    \theta_2 = \frac{\pi}{2}\pmod{\pi}\\
        \theta_3=\theta_1+\pi \pmod{2\pi}\\
        \theta_4 = 0 \pmod{2\pi}
    \end{array}
    \right.
\end{equation}
which provides two one-degree-of-freedom subsets (degree of freedom $\theta_1 \in \mathbb{R}$): \begin{itemize}
    \item 
$\mathcal{S}_{11}=\{(
    1,e^{i\theta_1},i,-e^{i\theta_1},1,-e^{i\theta_1},i,e^{i\theta_1}
)\}$ 
\item $\mathcal{S}_{12}=\{(
    1,e^{i\theta_1},-i,-e^{i\theta_1},1,-e^{i\theta_1},-i,e^{i\theta_1}
\}$.
\end{itemize}
We observe that $\mathcal{S}_{12}$ corresponds to the conjugate version of the set $\mathcal{S}_{11}$ which is consistent with the CAZAC properties. Moreover, one can note that the solutions found belong to the set $\mathcal{P}$. Indeed $\mathcal{S}_{11}$ corresponds to the second row in Table~\ref{tab:PP}.

Another less trivial solution satisfies $1+\cos(\theta_{4})+2\cos(\theta_3-\theta_1)=0$ with $\theta_4 \neq 0 \pmod{2\pi}$. After fancy calculations, we obtain \begin{itemize}
    \item $\mathcal{S}_{13}=\{(
    1,e^{i\frac{\nu+\rho}{2}},e^{i\gamma},e^{i\frac{\nu-\rho}{2}},e^{i\phi},e^{i\frac{\nu-\rho}{2}},e^{i\gamma},e^{i\frac{\nu+\rho}{2}}
)\}.$
\end{itemize}
with
\begin{equation}
    \left\{
    \begin{array}{lll}
    \chi &=& \sqrt{-2+2\sqrt{2}}\\
    \phi &=&2\mathrm{arcsin}\left(\chi\right)\\
    \gamma&=&- \mathrm{arccos}\left(-\frac{\chi^2}{2}\right)\\
    \rho &=& - \mathrm{arccos}\left(\frac{-1}{2}\left(1+\cos\left(\phi\right)\right)\right)\\
    \beta &=&\cos\left(\frac{1}{2}\left(\phi+\rho\right)\right)\\
    \tau &=&-\cos\left(\frac{\rho}{2}\right)\\
    \nu &=&2\left(\pi + \mathrm{arctan}\left(-\frac{\beta\cos\left(\frac{\phi}{2}\right)+\tau\cos\left(\gamma\right)}{\beta\sin\left(\frac{\phi}{2}\right)+\tau\sin\left(\gamma\right)}\right)\right).\end{array}
    \right.
\end{equation}

This type of sequence will be denoted by $\mathcal{C}^0_a$, with the uppercase 0 indicating a zero-degree-of-freedom sequence. The subscript $a$ referring to this type of equivalence class.   

\subsubsection{Second generic form}

The second form $\mathcal{F}_2$ has four parameters $\{\theta_1,\theta_2,\theta_4,\theta_5\}$ . Combining its structure to (\ref{eq:non_linear_system}) gives the non-linear system

\begin{equation}
    \left\{
    \begin{array}{lll}
    1+\cos(\theta_{40})+2\cos(\theta_{51})&=&0\\
    1+\cos(\theta_{20})+\cos(\theta_{42})+\cos(\theta_{51})&=&0\\
    e^{-i\theta_{50}}+e^{i\theta_{10}}+2\cos(\theta_{21})+2\cos(\theta_{25})&&\\
    +e^{i\theta_{41}}+e^{-i\theta_{45}}&=&0
    \end{array}
    \right.
\end{equation}

The first equation enables to find two first one-degree-of-freedom subsets of  solutions satisfying $\theta_4 =0\pmod{2\pi}$:
\begin{itemize}
\item $\mathcal{S}_{21}=\{\left(
    1,e^{i\theta_1},-i,e^{i\theta_1},1,-e^{i\theta_1},-i,-e^{i\theta_1}
\right)\}$,
    \item 
$\mathcal{S}_{22}=\{\left(
    1,e^{i\theta_1},i,e^{i\theta_1},1,-e^{i\theta_1},i,-e^{i\theta_1}
\right)\}$,
\end{itemize}
with the degree of freedom $\theta_1 \in \mathbb{R}$. $\mathcal{S}_{22}$ corresponds to the conjugate version of the set $\mathcal{S}_{21}$. Once again, those two types of solutions belong to the set $\mathcal{P}$ as $\mathcal{S}_{21}$ corresponds to the last row in Table~\ref{tab:PP}.

Then, if $\theta_4 \neq 0\pmod{2\pi}$, the non-linear system resolution yields zero-degree-of-freedom solutions grouped within the subset $\mathcal{S}_{23}$:
\begin{equation}
\small
   \begin{array}{rrrrrrrr}
   \{(1,&1,&z,&1,&-z,&-z,&z,&-z),\\
   (1,&1,&-1,&1,&-z,&-z,&-1,&-z),\\
     (1,&1,&z^*,&1,&-z^*,&-z^*,&z^*,&-z^*),\\
    (1,&1,&-1,&1,&-z^*,&-z^*,&-1,&-z^*),\\
   (1,&-z,&z,&-z,&-z,&1,&z,&1),\\
    (1,&-z^*,&z^*,&-z^*,&-z^*,&1,&z^*,&1),\\
    (1,&-z,&-1,&-z,&-z,&1,&-1,&1),\\
    (1,&-z^*,&-1,&-z^*,&-z^*,&1,&-1,&1)
     \}
           \end{array}
\end{equation}
where $\psi = \mathrm{arccos}(\frac{1}{3})$ and $z=e^{i\psi}$.
Four solutions among the eight zero-degree-of-freedom ones have the format $\bm{x}=\left(
    1,1,e^{i\theta_2},1,e^{i\theta_4},e^{i\theta_4},e^{i\theta_2},e^{i\theta_4}
\right)$ with $(\theta_2,\theta_4) \in \{(\psi,\psi+\pi),(\pi,\pi-\psi)\}$. We also observe that if $\bm{x}$ is in $\mathcal{S}_{23}$, then so is its conjugate version. 

The set of solutions will be referred to as $\mathcal{C}^0_b$. The use of the right combinations of translation, decimation, modulation and conjugation operations enables to show that the eight sequences are isomorphic and thus define a single equivalence class whose chosen representative in the remaining of the paper is ($(1,1,z,1,-z,-z,z,-z)$. 

\subsubsection{Third generic form}

The form $\mathcal{F}_3$ has five parameters $\{\theta_1,\theta_2,\theta_3,\theta_4,\theta_6\}$ and should verify the non-linear system 

\begin{equation}
    \left\{
    \begin{array}{l}
    e^{-i\theta_{30}}+e^{i\theta_{10}}+e^{i\theta_{21}}+e^{i\theta_{32}}+e^{i\theta_{43}}+e^{i\theta_{14}}
    \\
    +e^{i\theta_{61}}+e^{i\theta_{36}}=0\\
    e^{-i\theta_{60}}+e^{i\theta_{20}}+e^{i\theta_{42}}
    +e^{i\theta_{64}}+4\cos(\theta_{31})=0\\
    \cos(\theta_{40})+\cos(\theta_{26})+2=0
    \end{array}
    \right.
\end{equation}

The third equation imposes $\theta_4=\pi\pmod{2\pi}$ and $\theta_6=\theta_2+\pi\pmod{2\pi}$. Then, the full non-linear system resolution enables to find four one-degree-of-freedom subsets of solutions (degree of freedom $\theta_1 \in \mathbb{R}$):
\begin{itemize}
\item $\mathcal{S}_{31}=\{\left(
    1,e^{i\theta_1},1,-ie^{i\theta_1},-1,e^{i\theta_1},-1,-ie^{i\theta_1}
\right)\}$
\item $\mathcal{S}_{32}=\{\left(
    1,e^{i\theta_1},1,ie^{i\theta_1},-1,e^{i\theta_1},-1,ie^{i\theta_1}
\right)\}$ 
\item $\mathcal{S}_{33}=\{\left(
    1,e^{i\theta_1},-1,ie^{i\theta_1},-1,e^{i\theta_1},1,ie^{i\theta_1}
\right)\}$ 
\item $\mathcal{S}_{34}=\{\left(
    1,e^{i\theta_1},-1,-ie^{i\theta_1},-1,e^{i\theta_1},1,-ie^{i\theta_1}
\right)\}$.
\end{itemize}
where $\mathcal{S}_{32}$ (respectively $\mathcal{S}_{34}$) corresponds to the conjugate version of the set $\mathcal{S}_{31}$ (respectively $\mathcal{S}_{33}$).

Thus, this third form of sequences simply corresponds to sequences of set $\mathcal{P}$. Indeed $\mathcal{S}_{31}$ and $\mathcal{S}_{33}$ correspond to the first and third rows in Table~\ref{tab:PP} respectively.

\subsubsection{Fourth generic form}\label{sec:fourth_type}
  \begin{table*}
  \centering
  \caption{Fourth generic form: set $\mathcal{C}^0_c$. $\bm{s}_k~=~(0,0.5,s_k(2),\dots,s_k(7))$}
  \begin{tabular}{l|llllllll}
  &$s_k(0)$&$s_k(1)$&$s_k(2)$,&$s_k(3)$,&$s_k(4)$,&$s_k(5)$,&$s_k(6),$&$s_k(7)$\\
   \hline
 &   0&0.5&$a$&$4+b$&$3+c$&$7.5+c$&$1.5+b$&$6.5+a$\\
 $\mathcal{C}^0_c$ &0&0.5&$7.5+b$&$3+b-c$&$5.5-a+b$&$2-a+b$&$0.5+b-c$&$6+b$\\
  & 0&0.5&$7+c-a$&$6.5-a$&$1.5-a+b$&$6-a+b$&$4-a$&$5.5+c-a$\\
   &0&0.5&$6.5+b-c$&$7.5+a-c$&$5-c$&$1.5-c$,&$5+a-c$&$5+b-c$\\
   \hline
   &   0&0.5&$2-a$&$7-b$&$1-c$&$5.5-c$&$4.5-b$&$0.5-a$\\
    $\left({\mathcal{C}^0_c}\right)^\star$ &  0&0.5&$2.5-b$&$8-b+c$&$6.5-b+a$&$3-b+a$&$5.5-b+c$&$1-b$\\
    &   0&0.5&$3+a-c$&$4.5+a$&$2.5-b+a$,&$7-b+a$&$2+a$&$1.5+a-c$\\
   &0&0.5&$3.5-b+c$&$3.5+c-a$&$7+c$&$3.5+c$&$1+c-a$&$2-b+c$\\
      \hline
    \end{tabular} 
    \label{tab:fourth_form}
    \end{table*}
The solutions of the fourth form $\mathcal{F}_4$ are quite complex. Let
$(a,b,c)$ be the triplet solution of the set of non-linear equations:
\begin{equation*}
    \left\{
    \begin{array}{lll}
    \cos\left(\frac{\pi}{4}(a+3)\right)+\cos\left(\frac{\pi}{4}(a-b+\frac{1}{2})\right)&&\\
    +\cos\left(\frac{\pi}{4}(c-b+\frac{5}{2})\right)&=&0\\
    -\cos\left(\frac{\pi}{4}(a+3)\right)+\cos\left(\frac{\pi}{4}(b+\frac{5}{2})\right)&&\\
    -\sin\left(\frac{\pi}{4}(c-a)\right)+\cos\left(\frac{\pi}{4}(c-b+\frac{5}{2})\right)&=&0\\
    -\cos\left(\frac{\pi}{4}(c+5)\right)+\cos\left(\frac{\pi}{4}(b+\frac{5}{2})\right)&&\\+\sin\left(\frac{\pi}{4}(c-a)\right)&=&0
    \end{array}
    \right.
\end{equation*}
The application of the Newton method on a reformulation of the problem yields $12$ different CAZAC solutions among which $4$ are Popovic sequences with exact parameter values and the other are accurate approximations of two conjugate subsets depending on  $(a\;b\;c)=(0.1390361 \; 0.3487759 \; 0.0975818)$. From $a$, $b$ and $c$, it is possible to define 8 new CAZAC sequences defined by $\bm{x}_k = e^{i\frac{2\pi \bm{s}_k}{8}}$, $k = 1,2, \ldots, 7$ with $s_k(0)=0$, $s_k(1) = 0.5$ and the other components are  given in Table~\ref{tab:fourth_form}.

The two zero-degree-of-freedom subsets are referred to as $\mathcal{C}^0_c$ and $\left({\mathcal{C}^0_c}\right)^\star$. The corresponding CAZAC sequence subsets are congugate. The use of the right combinations of translation, decimation, modulation and conjugation operations enables to show that the sequences are isomorphic and thus define a single equivalence class whose chosen representative is  $\bm{x}_1=e^{i\frac{2\pi \bm{s}_1}{8}}$ ($\bm{s}_1=(0,0.5,a,4+b,3+c,7.5+c,1.5+b,6.5+a)$ in Table~\ref{tab:fourth_form}). Note that for this form, we have not been able to determine the exact formal expression of the solutions of the system of non-linear equations, and we hope that someone will be able to obtain them in the future. 

\subsubsection{Graphical classification of IPUC algorithm output}
The sequences obtained at the output of the IPUC algorithm are represented graphically in figure \ref{fig:graphe_phase} with a specific marker according to the set to which they belong ($\mathcal{P}$, $\mathcal{C}_a^0$, $\mathcal{C}_b^0$ and $\mathcal{C}_c^0$). Among the 10,000 CAZAC sequences, the proportions of each sequence are 7 \%, 30 \%, 30 \% and 33 \% for types $\mathcal{P}$, $\mathcal{C}_a^0$, $\mathcal{C}_b^0$ and $\mathcal{C}_c^0$, respectively. Note that other types of length $n=8$ CAZAC sequences may exist. In fact, we have no argument to evaluate whether the IPUC algorithm generates all possible types of CAZAC sequences, nor whether our exploration has been sufficiently thorough to present all possible types of CAZAC sequence potentially obtained by the IPUC algorithm. The question remains open.

\section{Conclusion}
In this paper we have dealt with the generation of CAZAC sequences. By exploiting Fourier transform conservation of the CAZAC property, we have proposed the IPUC algorithm which enables to generate CAZAC sequences of any length. Included within a simulated-annealing process, tailor-made sequences with additional features can be obtained. We illustrated this extension by generating sequences adapted for radar applications with optimized ratio between first and second lobes of the non-circular autocorrelation function. From a classification of the CAZAC sequences delivered by the IPUC algorithm and expoiting the polar-condition that CA sequences must satisfy to be ZAC, we have analytically derived four types of sequences in the case $n=8$. Among them, three cover the generic one-degree-of-freedom sequences found by Popovic completed by newly-identified zero-degree-of freedom equivalence classes. The newly-identified fourth type defines a zero-degree-of-freedom equivalent class whose chosen representative was written as a function of three parameters solution of a non-linear system. The exact formal expression remains to be found and the Newton algorithm applied on a reformulation of the problem enables us to determine an approximation. As future work, the method proposed here (IPUC algorithm combined with properties of CAZAC sequences) could be extended to classify CAZAC sequences of higher length which are not covered by existing families of the state-of-the-art (available for some specific values of $n$).

\section*{Acknowledgement}
\noindent This work has received funding from the FRench progrAm of IP Massification for Europe in xG (FRAME  xG) for the project C4xG.

\noindent The authors would like to thank Pr. Bertrand Banos of Université de Bretagne Sud for his precious help with solving the non-linear system in Section \ref{sec:fourth_type} to obtain $(a,b,c)$. He provided the authors with both the problem reformulation and the Newton algorithm to solve it.
%%%%%%%%%%%%%%%%%%%%%%%%%%%%%%%%%%%%%%%%%%%%%%%%%%%%%%%%%%%%%%%%%
\appendices

\section{Additional details about CAZAC sequences}\label{appendix:cazac_prop}
Let $\bm{x}_a$ be the version of $\bm{x}$ circularly shifted to the right by $a$ positions:
$\bm{x}_a=\left(
    x(n-a) , x(n-a+1) , \dots , x(0) , \dots ,x(n-a-1)
\right)$. We then define the matrix $\bm{\Omega}_x$ by \begin{eqnarray}
\bm{\Omega}_x=\begin{pmatrix}
    x(0)&x(1)&x(2)&\dots &x(n-1)\\
    x(n-1)&x(0)&x(1)&\dots &x(n-2)\\
    \vdots &\vdots &\vdots&\dots &\vdots\\
    x(1)&x(2)&x(3)&\dots &x(0)\\
\end{pmatrix}
\end{eqnarray} 
Then $\bm{R}_x=\bm{x} \bm{\Omega}_x^H$. Moreover DFT properties ensure that if $\bm{X}=\mathcal{F}\left(\bm{x}\right)$, then $\mathcal{F}\left(\bm{x}_a\right)(k)=e^{-i2\pi\frac{ka}{n}}X(k)$. One straightforward consequence is

\begin{eqnarray}
    \frac{1}{\sqrt{n}}\bm{F}^H\bm{\Omega}_x\bm{F}&=&\mathrm{diag}\left[\bm{X}\right]
\end{eqnarray}
where $\bm{F}=\frac{1}{\sqrt{n}}\begin{pmatrix}
    e^{-i2\pi \frac{k\ell}{n}}
\end{pmatrix}_{0\leq k,\ell \leq n}$ (Fourier matrix), $\bm{F}^H$ the conjugate transpose (Hermitian transpose) of $\bm{F}$, and $\mathrm{diag}\left[\bm{X}\right]$ stands for the diagonal matrix whose $k$-th diagonal element is $X(k)$.
Thus the relation
\begin{eqnarray}
    \bm{\Omega}_x\bm{\Omega}_x^H&=&n\bm{F}\mathrm{diag}\left[\bm{X}\odot \bm{X}^\star\right]\bm{F}^H
\end{eqnarray}
where $\odot$ stands for the Hadamard product (term-by-term vector multiplication). Similar reciprocal relationships exist between $\bm{\Omega}_{X}$ and $\bm{x}$.

As a consequence, the sequence $\bm{x}$ is ZAC if and only $\bm{\Omega}_x\bm{\Omega}_x^H$ is proportional to the identity matrix (that is to say $\bm{\Omega}_x$ is a Hadamard matrix \cite{Benedetto}). Then we immediately obtain that $\bm{x}$ is ZAC if and only if $\bm{X}$ is CA (and reciprocally). Moreover, for a CA sequence $\bm{x}$ defined by $x(k) = e^{i\theta(k)}$, $\bm{\Omega}_x\bm{\Omega}_x^H$ is proportional to the identity matrix if and only if the condition expressed in (\ref{eq:analytical_criterion}) is true.

\section{Some sequences obtained with Algorithm \ref{algo:projective_algorithm}}\label{annex:result}

In this section, the length-$n$ vector $\theta$ is represented by the length-$n$ vector $\textbf{s}$ defined as 
$\theta(k) = 2\pi s(k)/n$. We have stored in Table~\ref{tab:classifIPUC} some of the sequences delivered by the IPUC algorithm (we have not stored Popovic-like sequences). The type of sequence is then identified by applying to each generated sequence a combination of operations that preserve the
CAZAC property (see Table~\ref{table:transformations} in Section~\ref{sec:cazac_properties}).

{\small
\begin{table*}
    \centering
     \caption{Example of found sequences and associated class for $n=8$. The sequence expressed as $s$, with $s_i(0)=0$ and $\theta(k) =  2\pi s_i(k)/n$, $k=0,1,\ldots, 7$. The third column (function) shows how to transform each solution of a given equivalence class to one representative of the class thanks to a combination of the CAZAC-invariant transformations given in Table~\ref{table:transformations}. Note that the final rotation that allows to get the first coefficient equal to zero is omitted.} 
    \begin{tabular}{|c|c|c|c|} \hline
        Class &$(0,s_i(1),s_i(2),s_i(3),s_i(4),s_i(5),s_i(6),s_i(7))$& Function &Transformed sequence (Representative)\\ 
        
        \hline
        &$(0,4.346,1.456,2.566,2.912,6.566,1.456,0.346)$ &  $C_0 \circ M_2\circ D_5\circ T_0$ &$(0,0.566,5.456,6.346,2.912,6.346,5.456,0.566)$\\
        $\mathcal{C}^0_a$&$(0,7.890,0.544,5.890,4.000,6.110,7.456,4.110)$ &  $ C_0 \circ  M_3\circ D_3\circ T_6$ &$(0,0.566,5.456,6.346,2.912,6.346,5.456,0.566)$\\
        &$(0,1.346,3.456,1.566,6.912,7.566,7.456,3.346$) &   $C_0 \circ M_1 \circ D_5 \circ T_0$   &$(0,0.566,5.456,6.346,2.912,6.346,5.456,0.566)$\\
        &$(0,0.890,5.780,6.346,5.780,0.890,0.000,3.434$) &    $C_0 \circ M_4 \circ D_1 \circ T_1$   &$(0,0.566,5.456,6.346,2.912,6.346,5.456,0.566)$\\
        &$(0,2.566,1.456,4.346,2.912,0.346,1.456,6.566$) &   $C_0 \circ M_2 \circ D_7 \circ T_0$    &$(0,0.566,5.456,6.346,2.912,6.346,5.456,0.566)$\\
        &$(0,6.890,1.456,2.890,0.000,1.110,6.544,5.110$) &    $C_0 \circ M_2 \circ D_1 \circ T_2$   &$(0,0.566,5.456,6.346,2.912,6.346,5.456,0.566)$\\
        &$(0,6.566,4.000,5.110,2.220,3.654,6.220,5.110$) &    $C_0 \circ M_2 \circ D_7 \circ T_3$   &$(0,0.566,5.456,6.346,2.912,6.346,5.456,0.566)$\\
        &$(0,5.434,4.000,6.890,5.780,0.346,1.780,6.890$) &    $C_0 \circ M_2 \circ D_1 \circ T_7$  &$(0,0.566,5.456,6.346,2.912,6.346,5.456,0.566)$\\
        &$(0,3.890,3.456,1.890,4.000,2.110,4.544,0.110$) &    $C_0 \circ M_3 \circ D_7 \circ T_2$   &$(0,0.566,5.456,6.346,2.912,6.346,5.456,0.566)$\\
        &$(0,6.110,0.544,4.110,4.000,7.890,7.456,5.890$) &    $C_0 \circ M_3 \circ D_7 \circ T_6$   &$(0,0.566,5.456,6.346,2.912,6.346,5.456,0.566)$\\
        &$(0,1.110,1.456,5.110,8.000,6.890,6.544,2.890$) &    $C_0 \circ M_2 \circ D_5 \circ T_2$   &$(0,0.566,5.456,6.346,2.912,6.346,5.456,0.566)$\\
        &$(0,5.890,7.456,7.890,4.000,4.110,0.544,6.110$) &   $C_0 \circ M_3 \circ D_1 \circ T_2$    &$(0,0.566,5.456,6.346,2.912,6.346,5.456,0.566)$\\
        &$(0,7.110,2.220,1.654,2.220,7.110,8.000,4.566$) &  $C_0 \circ M_0 \circ D_1 \circ T_5$    &$(0,0.566,5.456,6.346,2.912,6.346,5.456,0.566)$\\
        &$(0,0.654,6.000,4.110,6.220,7.566,4.220,0.110$) &  $C_0 \circ M_3 \circ D_3 \circ T_7$    &$(0,0.566,5.456,6.346,2.912,6.346,5.456,0.566)$\\
        \hline
        &$(0,3.000,6.000,2.567,1.567,7.000,7.567,1.000)$ &   $C_1\circ M_1\circ D_1\circ T_5$   &$(0,0.000,1.567,0.000,5.567,5.567,1.567,5.567)$\\
        $\mathcal{C}^0_b$&$(0,3.000,0.433,7.433,6.433,1.433,2.000,3.433$) &  $C_1\circ M_1\circ D_5\circ T_7$ &$(0,0.000,1.567,0.000,5.567,5.567,1.567,5.567)$\\
       &$(0,2.433,6.433,2.433,2.433,0.000,6.433,8.000$) &   $C_1\circ M_0\circ D_5\circ T_0$   &$(0,0.000,1.567,0.000,5.567,5.567,1.567,5.567)$\\
       &$(0,0.000,4.000,2.433,0.000,2.433,6.433,0.000$) &$C_0 \circ M_4\circ D_5\circ T_2$   &$(0,0.000,1.567,0.000,5.567,5.567,1.567,5.567)$\\
       &$(0,2.000,5.567,6.000,5.567,7.567,5.567,3.567$) &   $C_0 \circ M_6\circ D_1\circ T_0$   &$(0,0.000,1.567,0.000,5.567,5.567,1.567,5.567)$\\
       &$(0,6.433,4.000,4.000,0.000,4.000,6.433,6.433$) &  $C_0 \circ M_0\circ D_1\circ T_2$    &$(0,0.000,1.567,0.000,5.567,5.567,1.567,5.567)$\\
       &$(0,1.567,1.567,1.567,5.567,4.000,1.567,4.000$) &  $C_0 \circ M_4\circ D_5\circ T_0$    &$(0,0.000,1.567,0.000,5.567,5.567,1.567,5.567)$\\
       \hline     
       &$(0,5.598,4.139,0.639,3.098,6.500,5.849,6.349$)&  $C_0 \circ M_2\circ D_5\circ T_0$   
       &$(0,0.500,0.139,4.349,3.098,7.598,1.849,6.639)$ \\
        $\mathcal{C}^0_c$       
        &$(0,2.500,3.849,1.251,5.710,4.210,4.751,4.349$)&   $C_1\circ M_0\circ D_3\circ T_1$   &$(0,0.500,0.139,4.349,3.098,7.598,1.849,6.639)$ \\
        &$(0,2.500,5.861,4.651,0.902,7.402,0.151,6.361$)&   $C_1 \circ M_5\circ D_1\circ T_0$   &$(0,0.500,0.139,4.349,3.098,7.598,1.849,6.639)$ \\
         &$(0,3.500,0.151,0.749,2.290,1.790,7.249,5.651$)&   $C_0 \circ M_6\circ D_3\circ T_1$   &$(0,0.500,0.139,4.349,3.098,7.598,1.849,6.639)$\\
         &$(0,2.651,1.249,0.790,6.290,1.749,6.151,6.500$)&   $C_0 \circ M_4\circ D_3\circ T_2$   &$(0,0.500,0.139,4.349,3.098,7.598,1.849,6.639)$ \\
        &$(0,2.749,3.249,1.500,2.290,7.651,4.151,7.790$)&  $C_0 \circ M_4\circ D_1\circ T_3$    &$(0,0.500,0.139,4.349,3.098,7.598,1.849,6.639)$\\
        &$(0,5.251,5.042,1.402,4.902,7.541,6.751,0.500$)&   $C_0 \circ M_5\circ D_1\circ T_5$   &$(0,0.500,0.139,4.349,3.098,7.598,1.849,6.639)$ \\
        &$(0,4.210,3.849,4.349,5.710,2.500,4.751,1.251$)&   $C_1 \circ M_0\circ D_7\circ T_5$   &$(0,0.500,0.139,4.349,3.098,7.598,1.849,6.639)$ \\
        &$(0,7.349,2.751,5.210,1.710,0.251,5.849,7.500$)&    $C_1 \circ M_6\circ D_3\circ T_2$   &$(0,0.500,0.139,4.349,3.098,7.598,1.849,6.639)$ \\
       \hline
    \end{tabular}   \label{tab:classifIPUC}
\end{table*}
}
\bibliographystyle{IEEEtran}
\bibliography{main.bib}

\begin{thebibliography}{10}
\providecommand{\url}[1]{#1}
\csname url@samestyle\endcsname
\providecommand{\newblock}{\relax}
\providecommand{\bibinfo}[2]{#2}
\providecommand{\BIBentrySTDinterwordspacing}{\spaceskip=0pt\relax}
\providecommand{\BIBentryALTinterwordstretchfactor}{4}
\providecommand{\BIBentryALTinterwordspacing}{\spaceskip=\fontdimen2\font plus
\BIBentryALTinterwordstretchfactor\fontdimen3\font minus \fontdimen4\font\relax}
\providecommand{\BIBforeignlanguage}[2]{{%
\expandafter\ifx\csname l@#1\endcsname\relax
\typeout{** WARNING: IEEEtran.bst: No hyphenation pattern has been}%
\typeout{** loaded for the language `#1'. Using the pattern for}%
\typeout{** the default language instead.}%
\else
\language=\csname l@#1\endcsname
\fi
#2}}
\providecommand{\BIBdecl}{\relax}
\BIBdecl

\bibitem{Benedetto}
J.~J. Benedetto, I.~Konstantinidis, and M.~Rangaswamy, ``{Phase-Coded Waveforms and Their Design},'' \emph{IEEE Signal Processing Magazine}, vol.~26, no.~1, pp. 22--31, 2009.

\bibitem{Benedetto2019}
J.~Benedetto, K.~Kosaian, and M.~Magsino, \emph{{CAZAC Sequences and Haagerup’s Characterization of Cyclic N-roots}}, 11 2019, pp. 1--43.

\bibitem{Chu_sequence}
D.~Chu, ``{Polyphase codes with good periodic correlation properties (Corresp.)},'' \emph{IEEE Transactions on Information Theory}, vol.~18, no.~4, pp. 531--532, 1972.

\bibitem{Chu_sequence2}
R.~Frank, S.~Zadoff, and R.~Heimiller, ``{Phase shift pulse codes with good periodic correlation properties (Corresp.)},'' \emph{IRE Transactions on Information Theory}, vol.~8, no.~6, pp. 381--382, 1962.

\bibitem{ZC_3}
\BIBentryALTinterwordspacing
D.~Gregoratti, X.~Arteaga, and J.~Broquetas, ``{Mathematical Properties of the Zadoff-Chu Sequences},'' \emph{CoRR}, vol. abs/2311.01035, 2023. [Online]. Available: \url{https://doi.org/10.48550/arXiv.2311.01035}
\BIBentrySTDinterwordspacing

\bibitem{chirp}
B.~Popovic, ``{Generalized chirp-like polyphase sequences with optimum correlation properties},'' \emph{IEEE Transactions on Information Theory}, vol.~38, no.~4, pp. 1406--1409, 1992.

\bibitem{Popovic2010}
B.~M. Popovic and O.~Mauritz, ``{Generalized Chirp-Like Sequences With Zero Correlation Zone},'' \emph{IEEE Transactions on Information Theory}, vol.~56, no.~6, pp. 2957--2960, 2010.

\bibitem{Kretschmer}
F.~Kretschmer and B.~Lewis, ``{Doppler Properties of Polyphase Coded Pulse Compression Waveforms},'' \emph{IEEE Transactions on Aerospace and Electronic Systems}, vol. AES-19, no.~4, pp. 521--531, 1983.

\bibitem{Wiener}
\BIBentryALTinterwordspacing
N.~Wiener, ``{Generalized harmonic analysis},'' \emph{Acta Mathematica}, vol.~55, no. none, pp. 117 -- 258, 1930. [Online]. Available: \url{https://doi.org/10.1007/BF02546511}
\BIBentrySTDinterwordspacing

\bibitem{Bjorck}
G.~Bj{\"o}rck, \emph{Functions of Modulus 1 on $Z_n$ Whose Fourier Transforms Have Constant Modulus, and ``Cyclic n-Roots''}.\hskip 1em plus 0.5em minus 0.4em\relax Dordrecht: Springer Netherlands, 1990, pp. 131--140.

\bibitem{10549571}
M.~Magsino and B.~Correll, ``{A Polar Optimization Search for CAZAC Sequences},'' in \emph{2024 IEEE Radar Conference (RadarConf24)}, 2024, pp. 1--6.

\end{thebibliography}
\end{document}